# A dynamic systems approach to harness the potential of social tipping

*Sibel Eker[a,b,*], Charlie Wilson[c,b,*], Niklas Höhne[d,e], Mark S. McCaffrey[f], Irene Monasterolo[g], Leila Niamir[b], Caroline Zimm[b]*

[a] Nijmegen School of Management, Radboud University, Netherlands
[b] International Institute for Applied Systems Analysis (IIASA), Austria
[c] Environmental Change Institute, University of Oxford, UK
[d] New Climate Institute, Germany
[e] Department of Environmental Sciences, Wageningen University, Netherlands
[f] UN Climate Change Community for Education, Communication and Outreach Stakeholders, Hungary
[g] EDHEC Business School, France
[*] Corresponding author

## Abstract

Social tipping points are promising levers to achieve net-zero greenhouse gas emission targets. They describe how social, political, economic or technological systems can move rapidly into a new state if cascading positive feedback mechanisms are triggered. Analysing the potential of social tipping for rapid decarbonization requires considering the inherent complexity of social systems. Here, we identify that existing scientific literature is inclined to a narrative-based account of social tipping, lacks a broad empirical framework and a multi-systems view. We subsequently outline a dynamic systems approach that entails (i) a systems outlook involving interconnected feedback mechanisms alongside cross-system and cross-scale interactions, and including a socioeconomic and environmental injustice perspective (ii) directed data collection efforts to provide empirical evidence for and monitor social tipping dynamics, (iii) global, integrated, descriptive modelling to project future dynamics and provide ex-ante evidence for interventions. Research on social tipping must be accordingly solidified for climate policy relevance.

## Introduction

The urgency for rapid and sustained reductions in greenhouse gas (GHG) emissions has drawn the attention of scientific and policy debates to social tipping points, which can trigger accelerated climate action through cascading effects in societies, institutions, and economic systems once a critical threshold is crossed. Therefore, social (or positive) tipping points[1,2] have gained wide attention as a high-leverage opportunity to counter-act upon high-risk climate tipping points[3] and to use limited policy resources most efficiently[4].

Social tipping points describe how social, political, economic or technological systems can move rapidly into a new system state or functioning[2]. The term often refers to nonlinear state change, without a clear distinction from similar phenomena, such as regime shift and critical transition[5]. A growing scientific literature, therefore, develops a definition and the theory of social tipping mechanisms, either harnessing an analogy to climate tipping mechanisms[1,6], or from a sociotechnical transitions perspective[2]. For instance, Milkoreit et al. [7] seeks a common definition by comprehensively surveying literature trends with various keywords related to social tipping. They find a rising publication count from the mid-2000s, dominated by disciplines of socio-ecological systems, climate change, and economics. Their content analysis of tipping point definitions emphasizes positive feedback structures as the core driver of nonlinear transitions between multiple stable states with limited reversibility, as well as multi-scale processes and cascading effects between systems.

In addition to alternative stable states, nonlinearity, positive feedbacks, limited reversibility as the four key attributes of tipping points, 'social' tipping points are characterized by desirability and intentionality in support of decarbonization and sustainability[5]. Furthermore, social systems involve more complex sets of interacting drivers and mechanisms, and do not have a single control variable [6]. These complexities mean single points or critical thresholds are difficult to isolate in social systems, therefore referring to processes and dynamics instead of the term social tipping point is more applicable[8].

The existing literature on social tipping dynamics, however, is missing a practical framework that embeds conceptual and empirical aspects of social tipping processes in order to inform tailored decisions, hence often exhibits an overuse of the term tipping point[5]. Therefore, the large potential beneficial impact of social tipping might be jeopardized by a weak analytical understanding due to the current limited and biased methods.

Here, we unpack the challenges that impede a strong analytical understanding of social tipping, and then propose a dynamic systems approach to tackle them. This dynamic systems approach aims to address the scientific purpose of a foundational understanding of system dynamics of social tipping, and the instrumental purpose of identifying effective interventions. It integrates (i) a systemic outlook on ST mechanisms that focuses not only on reinforcing but also impeding feedback mechanisms, as well as cascading effects across different subsystems, (ii) longitudinal data requirements and harmonization for empirical evidence and monitoring the effectiveness of interventions (iii) dynamic simulation modelling to explore the collective and cascading behaviour of feedback mechanisms and to create ex-ante evidence for effective ST interventions. In the following sections, we first summarize the current state of knowledge and debate on social tipping dynamics, then delineate the main challenges and outline the dynamic system approach we propose.

## What we know so far about social tipping

### *Social systems in which tipping can occur*

Several social systems can exhibit tipping dynamics. For instance, based on expert elicitation and literature review, Otto *et al.* [1] have identified six 'social tipping elements', that is, social, political, economic or technological systems in which tipping processes towards rapid decarbonization can occur. These are, as depicted in Figure 1, (i) energy production and storage, where subsidy programs and decentralized production can trigger rapid decarbonization; (ii) financial markets, where divestment from fossil fuels can rapidly reinforce investors' belief in the risks of carbon-intensive assets; (iii) education, where climate change coverage in school curricula can trigger sustained widespread engagement in climate action; (iv) norms and values, where advocacy by a few thought leaders can lead to a large population recognising anti-fossil fuel values; (v) urban infrastructure, where choosing clean technologies can trigger both cost reductions and consumer interest in pro-environmental choices, and (vi) information feedbacks, where disclosure of emission information on consumer products can trigger rapid behavioural change. Sharpe and Lenton [9] discuss the adoption of new technologies such as EVs and solar photovoltaics as specific examples related to energy and urban infrastructure. Farmer et al. [10] add institutional structures, e.g. UK Climate Change Act, since they can shape long-term and consistent climate policies. Taylor and Rising [11] focus on agriculture and demonstrate the presence of an economic positive tipping point beyond which the agricultural land use intensity starts declining.

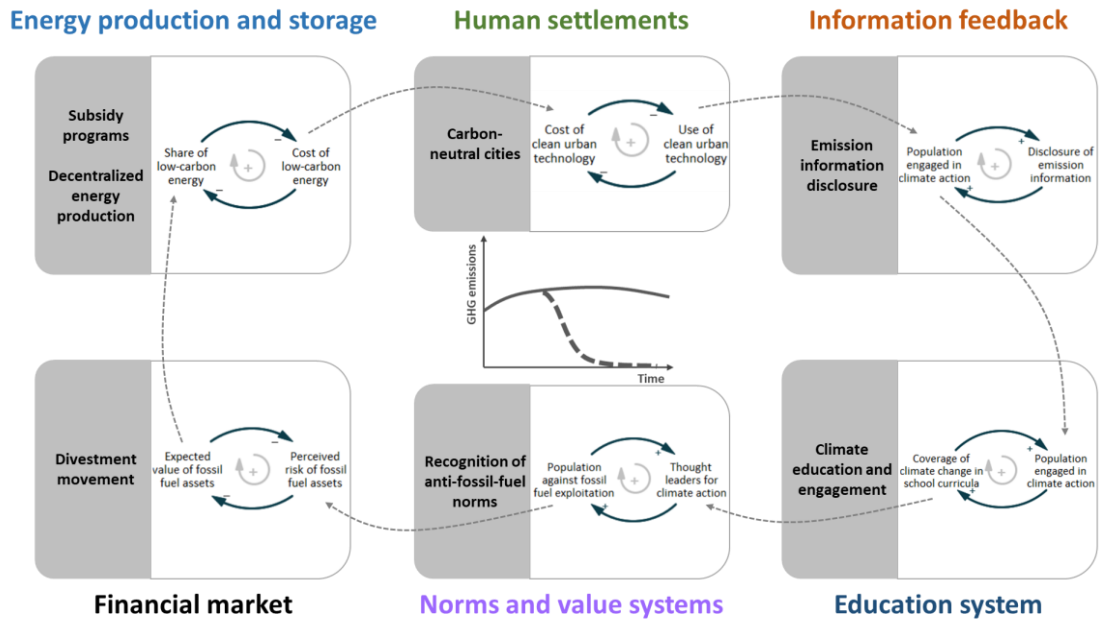

*Figure 1: Stylized depiction of six social tipping elements identified by Otto et al. (2020).* Social tipping elements (STE) refer to social systems in which tipping dynamics towards rapid decarbonization can be observed due to the annotated key positive feedback loops. The interventions that can trigger tipping dynamics in each elements are noted in grey. Besides the feedbacks within them, STEs have interconnections that can create cascading effects.

One of the biggest promises of ST dynamics is the cascading effects through interactions between the systems. For instance, Otto *et al.* [1] argue that more emphasis on climate change in the education system can lead to wider advocacy activities that trigger norm and value shifts while creating a higher sensitivity to carbon-emission disclosures on consumer products. Stadelmann-Steffen et al. [12] exemplify cross-system interactions with the historical phaseout of ozone-depleting chemicals. They consider Montreal Protocol, non-CFC substitutes and public concerns over UV radiation and skin cancer as interacting political, technological and behavioural tipping elements, respectively. Another example is provided by Pascual et al. [13] who identify the opportunities for positive tipping that emerge from the interactions between biodiversity, climate and society. Simulation results of Moore et al. [14] show a tipping behaviour in projected global carbon emissions resulting from cascading positive feedbacks through individual action, social conformity, climate policy and technological learning.

Besides cross-system interactions, cross-scale interactions can also trigger tipping dynamics as they result in contagion from individuals or organizations at the micro level to meso-level communities and macro-level countries and the world. For instance, renewable power and EV policies in a handful of frontrunner countries have been shown to accelerate the transition on a global scale across countries and sectors[9,15,16]. Similarly, a single schoolchild's protest has led to a global Fridays for Future movement, and through interconnections with other systems such as policy, it could create ST dynamics[12]. Interventions at the meso-level of communities (10000 - 100000 people) are identified to have maximum leveraging effect for rapid decarbonization[17], due to cross-scale interactions and pedagogy for agency[18].

### *Social tipping interventions*

ST interventions are active changes made to social systems in order to trigger or activate tipping processes, including those through cascading effects[1]. Such interventions can be 'kicks' that push the system onto a new trajectory without changing underlying structure but by triggering the loops (e.g., financial disclosure), or 'shifts' that change the system rules (e.g., institutional structures such as UK Climate Change Act)[10]. Not every climate change mitigation strategy, measure, action or policy can be considered a tipping intervention, unless they trigger or create relevant feedback loops underlying tipping dynamics.

National policies such as targeted investments, pricing policies, incentives, and regulations are considered ST interventions focused on feedbacks in specific systems[9,19]. Such interventions can also be triggered by civil society and create the constituency for government-led interventions[6,20] through cross-system interactions. For instance, behavioural interventions like communicating changes in social norms can accelerate demand-side mitigation, and positive spillovers can lead to tipping dynamics within or across consumption domains[21]. Moreover, reliance on market-based incentives, such as tax credits, may perpetuate wealth inequalities, weakening community empowerment and engagement, and acceptability of interventions. Therefore, ST interventions should be distinguished as those that can activate positive feedback mechanisms to trigger cascading dynamics across scales and systems.

### *Data availability and modelling*

Current scientific literature shows an inclination towards narrative-based presentation of potential social tipping dynamics, where empirical evidence is either in a limited context, or expert elicited and not observational . For instance, EV adoption is described as an example of cross-scale tipping dynamics in a narrative form[9], or possible tipping dynamics of coal phase-out in China is described based on an actor-objective-context framework[22].

Monitoring tipping processes is a data-intensive yet crucial activity to track if tipping threshold is approached or exceeded. . For instance, the transformation seismograph of New Climate Institute tracks indicators of tipping processes in power and transport systems[23]. Climate Action Tracker [19] monitors energy system indicators such as cost parity between renewable electricity generation and fossil-fuel assets. Systems Change Lab's data dashboard adds industry and finance indicators to these monitoring activities[24]. Similarly, Climate Watch monitors the policy system based on the records of countries that enhance their Nationally Determined Contributions (NDCs) or have net zero pledges in their law, policy documents or political pledges[25].

Quantitative simulations compile empirical evidence on individual systemic relationships from selected literature, market data or surveys, project the emerging long-term dynamics and demonstrate the conditions for tipping behaviour to occur. Existing evidence from simulation studies, though, remains limited to specific single systems, such as dietary change[26,27], global spread of urban innovations[28], urban cycling[29], or ground-water management[30]. The stylized global model of Moore *et al.* [14] notably combines multiple systems, from public opinion to individual technology adoption, climate policy and endogenous technological change, and show that individual action triggers a cascade of positive feedback processes through technological learning and social conformity for climate policy support.

## Challenges and knowledge gaps in the analysis of social tipping

To identify how feedbacks, multiple systems, cascading effects between them and evidence for social tipping dynamics are characterized, we scan the recent social tipping literature and find that (Figure 2, Supplementary Table 1) there is a clear weighting towards:

(i) single systems or scales where ST dynamics can occur, such as adoption of electric vehicle (EV) technology in the transport system, instead of multiple or connected systems (e.g. energy, finance, social norms, education) and scales (e.g. community, national, global);

(ii) a focus only on positive feedback mechanisms that can create the tipping dynamics sought by a particular perspective or agenda, but omit the tightly related negative feedback loops or undesired positive ones;

(iii) narrative or qualitative presentation of evidence for the account of social tipping (ST) dynamics, where the discussion remains mostly theoretical with empirical evidence obtained from selected literature.

Additionally, we observe that many case studies or empirical evidence are obtained from the Global North where different circumstances of the Global South are overlooked. There is an inherent degree of relativity in what is positive or negative, since a positive tipping outcome for one population may be viewed as disastrous for another.

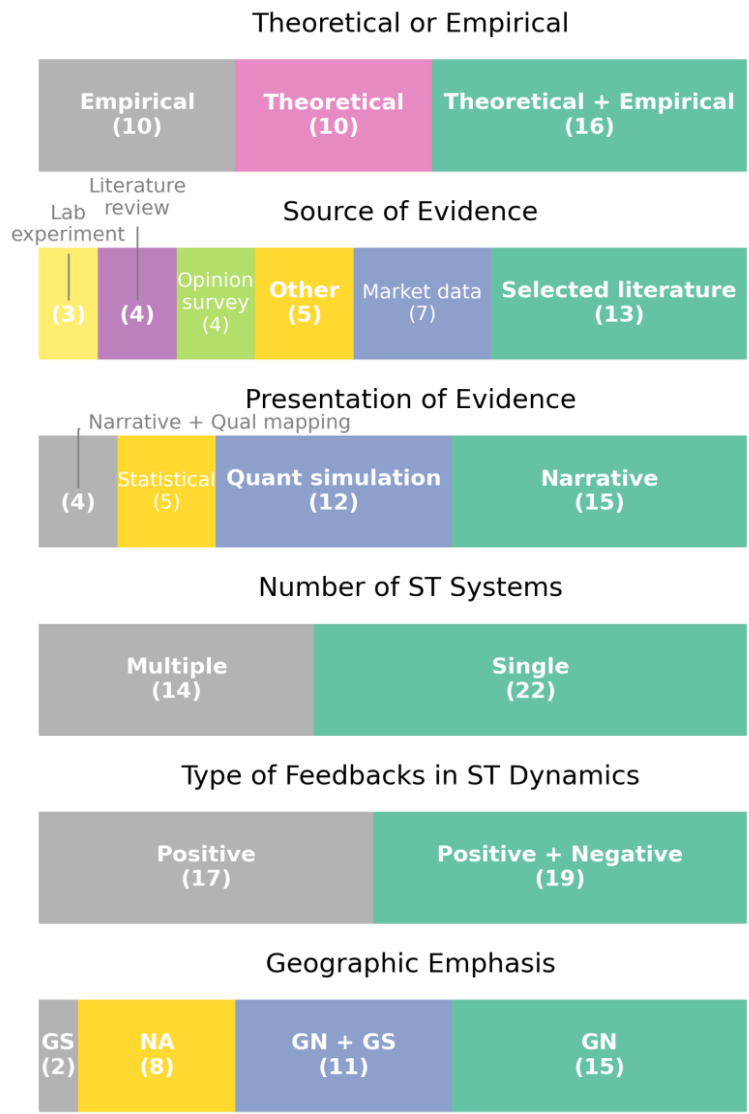

*Figure 2: Categorization of the emerging social tipping literature.* The publication data were retrieved from a search on Scopus database in October 2022 with search terms *"social tipping" OR "positive tipping" OR "sensitive intervention points" OR "socio-ecological tipping" OR "socio-economic tipping"* in article titles, abstract and keywords. We added five more articles identified during an expert workshop[8] to the resulting 59, and after screening for relevance, we categorized the remaining 36 articles. The rows of the figure refer to this categorization in terms of whether they provide empirical evidence or remain at a theoretical level, what source of evidence they use for tipping dynamics and how they present this evidence, whether they consider single or multiple social tipping (ST) systems, whether they focus on positive or negative feedbacks, and whether their geographic emphasis is on Global North or South. The numbers in parentheses refer to the number of publications in each category. 'NA' in the *Geographic Emphasis* row refers to the publications in which geographic coverage is not specified. 'Other' in the *Source of Evidence* row includes parametric evaluations or studies based on expert elicitation and selected literature reviews. The articles and full categorization can be seen in ***Supplementary Table 1.***

### *Focusing on single systems*

One of the biggest promises of ST dynamics is the cascading effects through interactions between the systems, yet these interconnections are sparsely examined as the single-system view in the existing literature shows. A single system focus without considering cross-system and cross-scale interactions, negative feedbacks, socioeconomic and geographic differences limits the scope and relevance of the intervention assessments. As sustainability transitions expert Frank Geels emphasizes[31], unlike relatively well-defined climate tipping points, analysing ST points requires taking the inherent complexity of social systems into account and all the efforts leading up to the tipping point. Therefore, the potential of ST interventions can be assessed with a more comprehensive approach aligned with social systems.

### *Focusing only on positive feedback loops*

The core driving mechanism of ST dynamics are positive feedback loops, hence most interventions proposed in the existing studies target those (Figure 2). Intervention outcomes are uncertain, though, due to the interactions between reinforcing (positive) and balancing (negative) feedback loops. Social movement interventions can trigger positive feedback loops of norm and value changes, for instance, yet they also cause value polarisation as a countervailing process. Efforts to stop specific lithium mines or wind farm projects and pipelines by local activists often runs counter to industrial scale climate policy ambitions. Protest movements against both fossil and low-carbon energy projects have stopped, suspended, or slowed new developments, but have also led to violence, with 10% of 649 cases analysed involving assassination of activists[32]. Polarisation also leads to a loss of diversity in opinion, ideas and solutions[33,34], undermining system resilience and jeopardizing the promise of interacting positive feedbacks for accelerated climate action. Such unintended consequences are also at the core of 'just transitions' research which addresses the coupled relationship between carbon emissions and ever-increasing wealth and energy inequalities, and highlights the need for a precautionary and holistic perspective on tipping and its justice implications[35]. Therefore, formulation of effective interventions can benefit from considering the role of negative feedbacks to avoid resistance and unintended consequences.

### *Lack of observational data and model-based studies as empirical evidence*

The empirical evidence underlying the theoretical, narrative-based discussion on social tipping often comes from selected, domain-specific literature. A few studies statistically show historical tipping dynamics based on large-scale data, such as European Social Survey[36] or gridded land use data[11]. A few lab experiments confirm the presence of tipping dynamics created by social conformity[37-39] where adoption of a new norm by the 25-40% of the population (critical mass) triggers further contagion. Even fewer field trials demonstrate the role of critical mass and information feedbacks in a real-life setting[40]. Such contextual and methodological limitations of empirical evidence cascades into the modelling studies that consolidate available data. Modelling studies are based mostly on a Global North perspective (Supplementary Table 1) and reflects neither the needs of future global consumers nor the complexity that contribute to socioeconomic and environmental inequality particularly in the Global South. Monitoring systems that aggregate national and global data are useful in tracking observed developments, and they can be expanded to social systems to include behaviour, norm and value changes with carefully selected metrics that indicate tipping dynamics and for which data can be collected.

The uptake of proposed tipping interventions by policymakers and stakeholders requires clear empirical evidence on their effectiveness. Therefore, geographically and contextually more comprehensive statistical, experimental and modelling studies are needed to establish such clarity of evidence.

# Dynamic systems approach to social tipping

To address the gaps in the conceptualization and assessment of ST points and interventions, we introduce a three pillared dynamic systems approach with examples developed in an expert workshop that focused on participatory modelling of key social tipping processes[8]. The three pillars refer to delineations of system structures, quantifying and monitoring tipping dynamics, and dynamic modelling to consolidate available empirical knowledge and evaluate potential interventions.

## *Systems outlook*

Understanding potential tipping dynamics for rapid decarbonization can be enhanced by delineating the underlying system structure based on three principles:

> i. *Characterize and map the feedback mechanisms in each social tipping system, by taking potential barriers to positive tipping dynamics into account.*

ST processes described in many existing studies depict the mental models of experts from physical climate science or social sciences such as transition studies based on sector-specific historical behaviour. These mental models often focus on the critical threshold of a tipping process, describe a unidirectional impact from interventions to outcomes, and do not always explicate closed chains of relationships (feedback loops). Delineating the feedback mechanisms, however, can lead to a better understanding of the eventual dynamic system behaviour.

ST dynamics are expected to occur as a result of positive (reinforcing) feedback mechanisms[41,42] that amplify a change in the same direction through a loop of system elements. Existing conceptualizations of ST processes emphasize such feedbacks that positively affect decarbonization and overlook the negative ones (Figure 2). Even though tipping dynamics are characterized by reinforcing feedbacks, dynamic systems are characterized by multiplicity of coupled negative and positive feedbacks[43]. Therefore, considering negative effects and feedbacks can provide a more balanced estimate of the tipping potential and help avoid unintended consequences of interventions. For instance, rapid divestment from fossil fuel assets is considered a financial tipping intervention[1], yet it can lead to financial instability and adverse distributional consequences that can undermine system functioning[8]. Diffusion of ethical values against fossil fuel exploitation through social conformity is another key social tipping process[1]. This reinforcing loop of social conformity, though, is counter-acted in reality by the feedback mechanisms of polarization and industry resistance, which might impede the tipping potential of norm changes. Moreover, what may be considered a positive tipping in the Global North, e.g. rapid and large-scale decarbonization, may trigger unintended negative consequences in the Global South and other marginalized regions, such as the closing down of wealth-generating markets and export opportunities, if not planned for.

Multiple methods can be employed in combination to delineate the feedback mechanisms underlying tipping dynamics. For instance, participatory systems mapping methods either based on causal loop diagrams[44] or fuzzy cognitive maps[45] can elicit and align expert views. Qualitative or semi-quantitative models co-developed using these participatory methods can be complemented by literature reviews for quantitative empirical evidence. Box I and Figure 3 exemplify coupled feedback loops delineated in a participatory modelling workshop and supported by empirical studies listed in Supplementary Table 2.

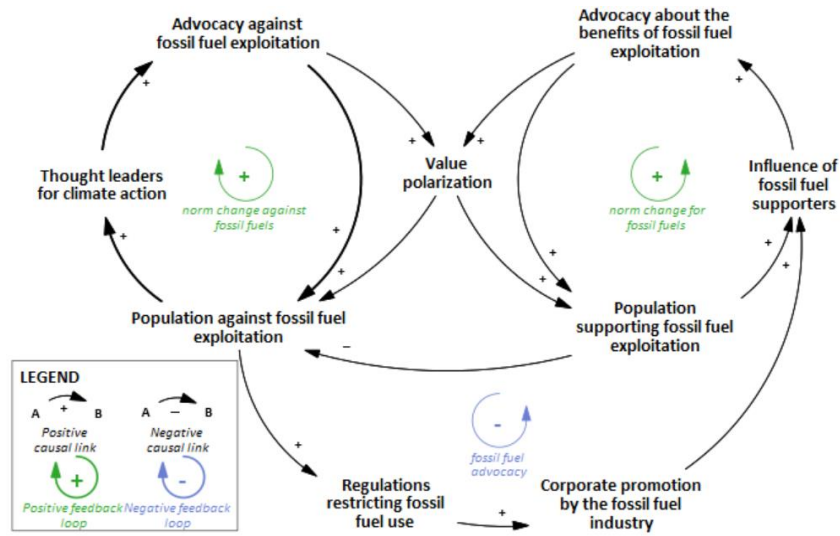

*Figure 3: Main feedback loops underlying the social tipping dynamics in the norms and values system,* derived from expert elicitation and supported by the empirical studies listed in Supplementary Table 2. A positive causal link implies that a change in variable A changes variable B in the same direction, whereas a negative link implies a change in the opposite direction. A positive feedback loop refers to a closed chain of relationships that includes an even number of negative links, and where a change in any element, either in the positive or negative direction, is reinforced through the loop. A negative feedback loop refers to a closed chain with an odd number of negative links, where a change is balanced through the loop.

---

**Box I: Multiple positive and negative feedback mechanisms governing norm and value changes**

Since social and moral norms are a key driver of human behaviour[46], shifting towards anti-fossil fuel norms is considered a key social tipping process for rapid decarbonization[1]. Advocacy against fossil fuel extraction even by a small group of thought leaders or influencers can stimulate the diffusion of pro-environmental values[10]. The feedback loop *norm change against fossil fuels* in Figure 3 depicts this reinforcing mechanism of diffusion: *Thought leaders* who advocate for anti-fossil-fuel norm changes can be individuals or organizations within civil society, international organizations, state leaders and subnational governments[47]. Their advocacy activities are empirically shown to influence public opinion and mobilization against fossil fuel exploitation, as exemplified by the individual influence of Bill McKibben[48] and Greta Thunberg[49], or the student activists mostly influenced by their leaders[50]. As the population against fossil fuel exploitation increases, more thought leaders or norm entrepreneurs emerge from different communities and newly created coalitions[51], closing the loop of diffusion.

In contrast, the reinforcing loop of *norm change for fossil fuels* acts as a primary impediment to anti-fossil fuel norm shifts, since it represents a value polarization cycle commonly observed in climate debate in multiple countries[52,53]. Recent lab experiments also show that identity and polarization are strong impediments to tipping dynamics in a broader context[39]. Pro-fossil fuel norms develop similarly to the anti-fossil fuel norms: *Population supporting fossil fuel exploitation* increases as the *advocacy about the benefits of fossil fuel exploitation* becomes prevalent, as exemplified by the strong relation between public opposition to one of the major

US climate policies and views of politicians and certain TV channels[54]. In return, political leaders adopt a polarizing language to appeal to the increasing fraction of population supporting their view[55], which enhance advocacy activities and make fossil fuel policies one of the most politically polarized issues especially in the US[52]. People who are exposed to opposing views stick to their own view more strongly[56], hence advocacy activities enhance value polarization and reinforce the norm change feedbacks on both sides. The amplifying effect of partisan identification on climate policy support among both for Republicans and Democrats in the US[57] exemplify the role of such feedbacks.

A balancing feedback mechanism that affects norm shifts is the *fossil fuel advocacy* loop. As the population against fossil fuel exploitation increases, the resulting social mobilization leads to policies that restrict fossil fuel extraction and use, as observed in many local and national settings so far[32,58]. *Regulations restricting fossil fuel use* is the main driver of *corporate promotion by the fossil fuel industry*[59], which enhances the pro-fossil fuel advocacy activities[60], and eventually reduces the *population against fossil fuel exploitation*. This feedback loop potentially dampens the growth of *population against fossil fuels*, hence the norm shifts. A similar balancing loop can be formulated due to the media coverage of climate change leading to higher pro-fossil fuel advertisements[59], often triggered by advocacy activities of influential thought leaders. The real-world example of fossil fuel resurgence following the war in Ukraine provides an opportunity to examine how these dynamics can play out on the world stage.

ii. *Identify and map the interactions across multiple systems in the conceptualization of social tipping processes.*

The analysis of social tipping processes tends to be system-specific, such as the diffusion of EVs in the transport sector[9]. However, many of these systems are strongly interconnected, as exemplified by the education-society links for shifting to anti-fossil fuel norms[1], or policy-technology-behaviour connections that tipped the phaseout of CFCs[12]. Growing empirical evidence supports the presence of interactions between public opinion, social norms, individual pro-environmental behaviour, climate policy, climate impacts (and their effects on opinion), and technological change[14,61]. The dynamic behaviour resulting from these cross-system interactions that potentially lead to additional feedback mechanisms might accelerate tipping dynamics and boost the effectiveness of interventions, or vice versa.

The examples of cross-system interactions provided so far are limited in scope, and further interconnections can be identified and analysed, for instance, between education, finance and energy systems. Estimating and validating the tipping potential of interventions can benefit from maintaining a feedback perspective in specifying these interconnections, rather than formulating them as linear cascading effects, and from the crucial support of empirical findings.

Participatory approaches with experts and stakeholders from different communities can facilitate interdisciplinary research needed to identify the existing and potential cross-system interactions. While providing quick access to well-informed mental models and available empirical evidence, participatory research can also steer new empirical research for quantifying cross-system interactions. Such participatory approaches themselves can contribute to social tipping through their impact on social movements[62]. Box II and Figure 4Figure 3 exemplify interactions between energy, finance, urban infrastructure, policy and society delineated in a participatory modelling workshop and supported by empirical studies listed in Supplementary Table 3.

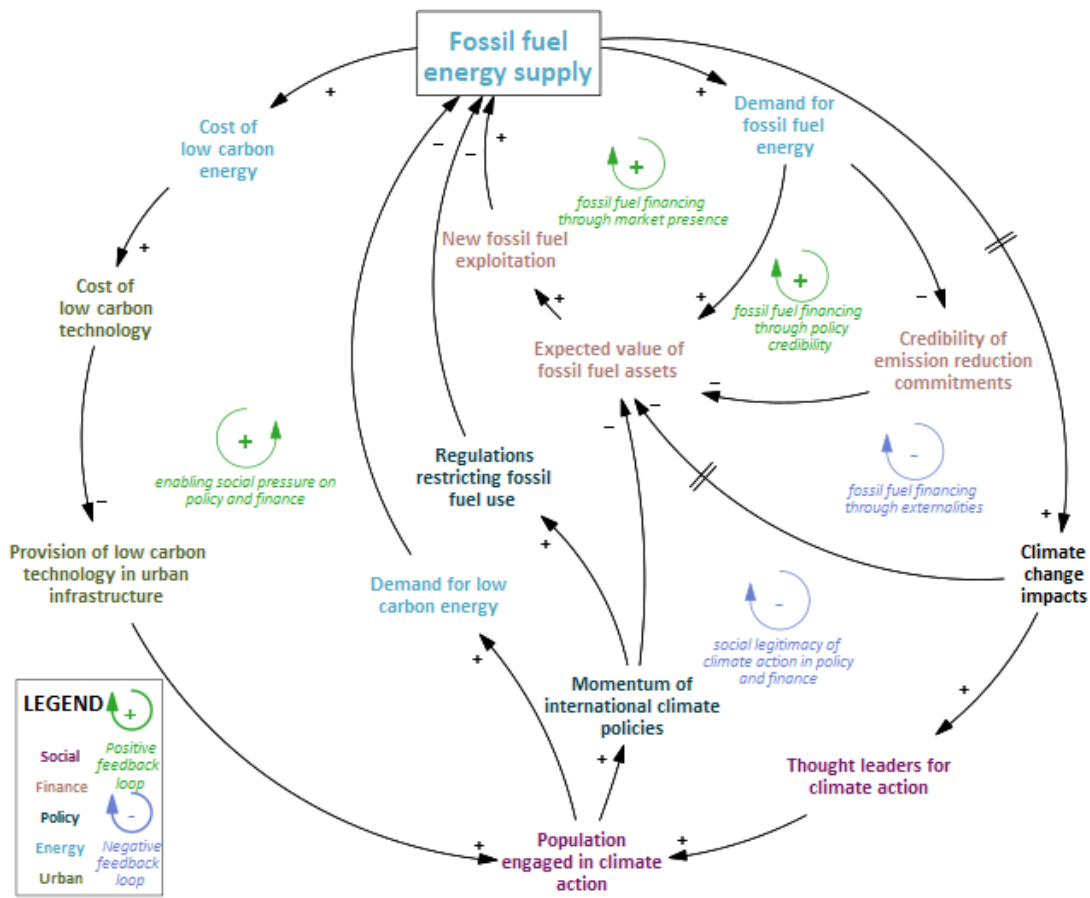

*Figure 4: Main feedback loops resulting from the interactions between energy, finance, urban, social and policy systems*, derived from expert elicitation and supported by 27 empirical studies listed in Supplementary Table 3. Double lines on arrows indicate a delay in the relationship depicted by that arrow. (See the caption of *Figure 3* for explanation of the notation.)

> **Box II: Cross-system interactions**
>
> Energy, finance, policy, societal and urban infrastructure systems involve positive feedback mechanisms that can individually lead to social tipping dynamics[1,8]. They also interact with each other through linear cascading effects and wider feedback loops that can amplify or dampen the tipping dynamics. Figure 4 depicts those interactions which are based on an expert co-modelling workshop and the empirical studies listed in Supplementary Table 3.
>
> The loop *fossil fuel (FF) financing through market presence* shows a key coupling of finance and energy systems: The higher the *fossil fuel energy supply*, the higher the demand, leading to higher *expected value of FF assets*[63], hence more investment and higher FF energy supply[64] (and the reverse applies). This feedback loop is further reinforced by the *credibility of emission reduction commitments*. If investors trust climate policy announcement and introduction, they would revise their risk assessment for FF firms, leading to higher cost of capital for FF investments, lowering profitability, thus the FF asset value[65,66]. Credibility of commitments leads also to a lower cost of capital for renewable energy investments, further enabling decarbonization. Credibility of commitments is reduced by a continuing high demand for FFs but enhanced by the strength of climate policies itself[67]. The expected value of FF assets is also dependent on perceived *climate change impacts*[68], which creates the balancing feedback loop of *FF financing through externalities* as diminishing FF supply would reduce the climate impacts in the long term.

Another major driver of expected value of FF assets is the *momentum of international climate policies*. For instance, the Paris Agreement led to a significant reduction in the high-carbon stock values and an increase in the cost of borrowing[69]. International climate policies eventually reduce the FF supply through not only their financial impacts but also their direct impact on national *regulations restricting fossil fuel use*[70]. National policies such as carbon tax or emission trading focus on fossil fuel consumption, yet those restricting supply has gained momentum[71]. Their impact on global FF supply is yet to be achieved[72], as the location of such policies and FF extraction match[73].

The balancing loop *social legitimacy of climate action* depict the influence of social changes on the FF energy supply through finance and policy: *Population engaged in climate action* through direct mitigation behaviours such as energy saving or civic action enhances the momentum of international climate policies by putting pressure on negotiations and signalling the readiness for national policies[74]. Worsening *climate change impacts* increase the engagement in climate action either directly[61,75], or indirectly via *thought leaders*[76] who communicate the climate change causes and solutions. Climate impacts are dependent on *fossil fuel energy supply*, which can be traced back to the *momentum of climate policies*.

*Enabling social pressure* loop depicts the connection of urban infrastructure, energy, policy and finance systems: *Provision of low-carbon urban technology* can facilitate low-carbon behaviours such as reducing household waste and energy use[77] or cycling[78], increasing the population engaged in climate action and eventually lowering fossil fuel energy supply. This in return can reduce the *cost of low-carbon energy*, subsequently the *cost of low-carbon urban technologies*, resulting in further provision of low-carbon urban technology[79].

iii. *Identify and map the interactions across multiple scales in the conceptualization of contagion dynamics that lead to social tipping*

Social contagion among individuals is a strong feedback loop that triggers tipping dynamics[80]. Contagion can also be observed among and across communities, firms, authorities and nations[81], resembling fractals that replicate the same structure[18]. Acknowledgement of different scales of agency and their cross-scale interactions helps to overcome the fractal carbon trap[15], where the decision-making agency is attributed to a single actor or ideology (such as free-market solutions to social, economic and environmental problems) towards diverse, multilevel, catalytic action at different scales.

System conceptualization can explicate the scale of each tipping mechanism, such as individuals, multinational corporations or national governments, and identify the bi- or multi-lateral interactions between those scales. Box I exemplifies the contagion effects among individuals, and how these relate to firm-level actions and national policies. Such an explicit account of different scales of action and their interactions also helps formulating tipping interventions to fulfil dynamic needs of the society and capture arising opportunities, beyond achieving a static goal such as emission reductions[82].

## *Data gathering and harmonization*

Complementing the system conceptualization described above, dedicated data collection efforts are needed to move beyond specific, single system data from selected literature; to consolidate empirical evidence for conceptual feedback loops underlying tipping dynamics as exemplified in Box I and II; and to monitor the actual or potential effectiveness of interventions.

Data collection requires identifying the key indicators that can represent the dynamics created by coupled feedback mechanisms and interventions. For instance, cost parity between low-carbon and fossil-fuel energy supply combines the dynamics of technological learning and economies of scale feedbacks both from the low-carbon and fossil fuel energy sector. Box III and Figure 5 exemplify two monitoring variables identified in an expert workshop.

---

**Box III : Monitoring social tipping dynamics**

Monitoring social tipping dynamics requires operationalizing variables that can capture the cascading feedback dynamics in multiple systems. Below are two examples of such variables, with Figure 5 presenting a stylized potential trajectory of each variable.

**Number of systemically important companies calculating climate Value-at-Risk** is an indicator of climate risk perception in the financial markets, hence the perceived risk of fossil fuel assets. Systemically important companies can be defined as those which have more than $ 100 billion in assets. We estimate this variable to have increased increasingly in recent years, but the critical threshold is yet to be achieved in the next few years.

**Willingness to pay for climate action** can be used to monitor population engaged in climate action, for which data is already collected and used in some contextual studies. Willingness to pay is heavily dependent on income level and economic situation, hence expected to fluctuate over time. In the middle income group, willingness to pay is expected to have an increasing future trend, whereas it is estimated to be well below a critical threshold currently in the low-income group.

---

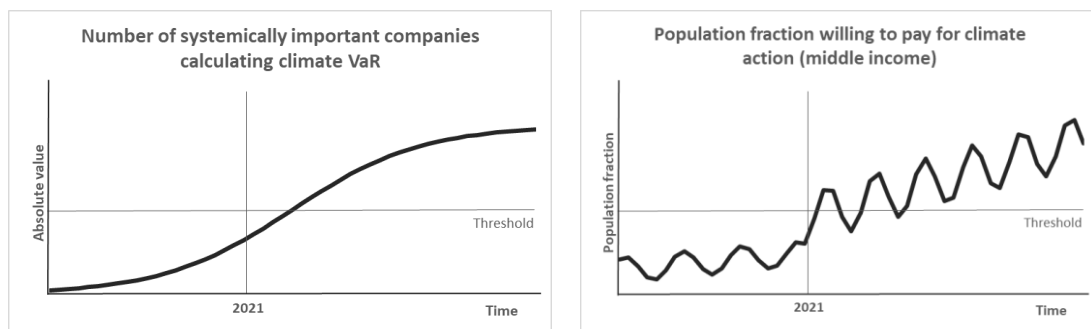

*Figure 5: Stylized trajectories of monitoring variables,* Number of systemically important companies calculating *climate Value-at-Risk* at left, *and Willingness to pay for climate action* at right.

Monitoring ST dynamics requires harmonizing different sources of time-series data on common time frames to enable detecting cascading cross-system changes. For instance, social media data can be used as a high-frequency, publicly available and low-cost global source[83] to monitor the norm and value changes in social systems, in combination with purposeful, lower frequency data such as World Values Survey[84]. Harmonizing this data on norm and value changes with records of other systems, such as international and national policy action, energy cost parities, technology adoption levels and financial flows, can help quantifying the bilateral cross-system relationships and monitoring their cascading effects towards tipping. Sharing these harmonized data on online platforms can facilitate further in-depth collaborative research within scientific communities, whereas public display can demonstrate the importance of rapid action.

### *Dynamic modelling*

Modelling is a key tool in analysing and navigating dynamic systems, helping understand how a system works, and bringing rigor to the analysis with an explicit formulation of ideas and assumptions, consolidation of data, and logical tracing of those formulation sequences. Models, either qualitative or quantitative, provide a future outlook by estimating how a variable is likely to evolve, diagnose what factors have the greatest leverage to change outcomes, and assist in ex-ante policy assessments. In the social tipping context, quantitative modelling is commonly used (Figure 2) to demonstrate the conditions under which tipping occurs, yet in stylized cases and mostly from a single system perspective.

Dynamic modelling can support the analysis of social tipping dynamics by embedding four key aspects. First, models should enable a systematic demarcation of interconnected feedback mechanisms within multiple systems and their cross-system and cross-scale interactions from micro to meso and macro level. Second, the models should be grounded in representative data and move towards quantitative realm for computational analyses of feedback dynamics. Quantifying social systems at a global level is challenging and aggregation in stylized representations is unavoidable. Still, quantitative methods aligned with the global data can provide actionable evidence for the long-term effectiveness of interventions, while qualitative and participatory approaches facilitate conceptualization and dissemination of such quantitative modelling. Third, to tackle the broad scope of multiple social systems and feedbacks, an iterative modelling approach can help, where broad system boundaries are narrowed down through empirical support and computational diagnostic analyses, and further research efforts are dedicated to those feedbacks that are shown to be more important. Fourth, interdisciplinary modelling can ensure policy relevance through either hard or soft coupling between models of tipping dynamics and the existing climate policy models.

A modelling discipline that has widely and influentially guided climate policy assessments is integrated assessment modelling (IAM)[85], yet considered limited in covering nonlinear social and behavioural processes[86,87]. Social tipping processes can be more suitably captured by an emerging group of simpler, aggregate IAMs developed with descriptive, rather than optimization-based, dynamic modelling methods. Current examples of such models include those developed with agent-based modelling (ABM) of different economic sectors[88,89], and those developed with system dynamics (SD) modelling based on aggregate representation of cross-sectorial feedbacks[90,91]. This emerging group of models that incorporate social tipping processes intersects with social climate models that focus on representing human and Earth system feedbacks[92].

These simple models represent nonlinear relationships and feedbacks, are more flexible to scope extensions compared to conventional IAMs, can be more easily calibrated to emerging data from the monitoring systems, and facilitate computational analyses with large numbers of simulations. ABMs are powerful in modelling social contagion, and often used with threshold models[93] where the decision of a given actor is formulated conditional on the number of others who make that decision[80], hence also used in modelling social tipping dynamics[94]. Recent evidence, though, shows that threshold models may not represent the nonlinearities of real life[36], and ABMs are often prone to overcomplexity and are weaker in terms of statistical validation of results, making sometimes hard to identify policy recommendations. Stock-flow consistent (SFC) models[95] models merge desirable behavioural features of ABM with robust balance sheet accounting in which heterogeneous agents, sectors and their financial flows are represented as a network of interconnected balance sheets, allowing for tracing causal relations and validation of results, and contributing to overcome the limitations of ABMs.

SD modelling can suit exploring the global social tipping dynamics, primarily because interconnected feedbacks within and across multiple systems can be better represented in the aggregate and feedback-rich view in SD. ABMs require micro-level data for calibration and validation[96,97], and the computational requirements for micro modelling at global scale might hinder uncertainty analysis and interactive simulations[96,98]. Therefore, an aggregate modelling view can better suit the available data at a global level. Since complexity of micro-phenomena on a global scale impedes relating the model behaviour to the structure[97,99], SD models can also allow deriving cognitively grounded insights.

In previous global modelling studies, based on coupling social and behavioural feedback mechanisms with those of land use and climate dynamics, Eker et al. [27] showed that triggering social norm feedbacks at an early stage of diffusion is the most influential driver of widespread shifts to plant-based diets. Moore et al. [14] presented a prominent example of cross-system modelling found that low-emission trajectories consistent with Paris Agreement targets can emerge through positive tipping dynamics for which social conformity, technological learning, political responsiveness to public opinion, and cognitive biases in perception of climate impacts are the key. Similar modelling studies can cover additional high-leverage systems and connections, such as energy and finance, with more nuanced and policy-relevant representation of tipping elements. Quantification of these models with globally representative data[92], including those from Global South and disenfranchised populations of the Global North, can enable defining trajectories against which actual change is monitored, so that system structure and behaviour can be better understood. Subsequently, this better understanding and empirical grounding enhance the usefulness of models in analysing the effectiveness of social tipping interventions.

## Way forward

Social tipping points have gained wide attention in the scientific and policy debate as high-leverage and cost-efficient options to accelerate emission reductions. The growing scientific literature on social (positive) tipping points, though, is dominated by a narrative account of the social tipping dynamics, lacking a clear empirical basis, and by a focus on single technologies, systems, scales, and feedback mechanisms that originate from the Global North that often exclude a social justice perspective. Harnessing the promising potential of social tipping dynamics, though, requires wide-ranging and systematic analyses with multiple empirical methods and with a broad systems outlook that involves multiple systems, agency scales and interconnected feedback loops.

Here, we outlined a dynamic systems approach that involves a systems outlook with positive and negative feedback loops withing and across multiple systems and scales; concurrent data collection in multiple systems not only to provide empirical evidence for tipping dynamics but also to monitor them; dynamic simulation modelling to consolidate conceptual and empirical knowledge and for ex-ante analysis of tipping interventions. We argue that it is critical to use such a systems approach to better understand social tipping dynamics, to ensure climate policy relevance. This approach can help solidify the popularity of the social tipping concept in better-informed policies and practices.

# References

1. Otto, I.M., Donges, J.F., Cremades, R., Bhowmik, A., Hewitt, R.J., Lucht, W., Rockström, J., Allerberger, F., McCaffrey, M., Doe, S.S.P., et al. (2020). Social tipping dynamics for stabilizing Earth's climate by 2050. Proceedings of the National Academy of Sciences *117*, 2354-2365. 10.1073/pnas.1900577117.
2. Tàbara, D.J., Frantzeskaki, N., Hölscher, K., Pedde, S., Kok, K., Lamperti, F., Christensen, J.H., Jäger, J., and Berry, P. (2018). Positive tipping points in a rapidly warming world. Current Opinion in Environmental Sustainability *31*, 120-129. https://doi.org/10.1016/j.cosust.2018.01.012.
3. Armstrong McKay, D.I., Staal, A., Abrams, J.F., Winkelmann, R., Sakschewski, B., Loriani, S., Fetzer, I., Cornell, S.E., Rockström, J., and Lenton, T.M. (2022). Exceeding 1.5C global warming could trigger multiple climate tipping points. Science *377*, eabn7950. doi:10.1126/science.abn7950.
4. Efferson, C., Vogt, S., and Fehr, E. (2020). The promise and the peril of using social influence to reverse harmful traditions. Nature Human Behaviour *4*, 55-68. 10.1038/s41562-019-0768-2.
5. Milkoreit, M. (2022). Social tipping points everywhere?—Patterns and risks of overuse. WIREs Climate Change *n/a*, e813. https://doi.org/10.1002/wcc.813.
6. Winkelmann, R., Donges, J.F., Smith, E.K., Milkoreit, M., Eder, C., Heitzig, J., Katsanidou, A., Wiedermann, M., Wunderling, N., and Lenton, T.M. (2022). Social tipping processes towards climate action: A conceptual framework. Ecological Economics *192*, 107242. 10.1016/j.ecolecon.2021.107242.
7. Milkoreit, M., Hodbod, J., Baggio, J., Benessaiah, K., Calderón-Contreras, R., Donges, J.F., Mathias, J.-D., Rocha, J.C., Schoon, M., and Werners, S.E. (2018). Defining tipping points for social-ecological systems scholarship—an interdisciplinary literature review. Environmental Research Letters *13*, 033005. 10.1088/1748-9326/aaaa75.
8. Eker, S., and Wilson, C. (2022). System Dynamics of Social Tipping Processes. International Institute for Applied Systems Analysis (IIASA). https://pure.iiasa.ac.at/id/eprint/17955/.
9. Sharpe, S., and Lenton, T.M. (2021). Upward-scaling tipping cascades to meet climate goals: plausible grounds for hope. Climate Policy *21*, 421-433. 10.1080/14693062.2020.1870097.
10. Farmer, J.D., Hepburn, C., Ives, M.C., Hale, T., Wetzer, T., Mealy, P., Rafaty, R., Srivastav, S., and Way, R. (2019). Sensitive intervention points in the post-carbon transition. Science *364*, 132-134. doi:10.1126/science.aaw7287.
11. Taylor, C.A., and Rising, J. (2021). Tipping point dynamics in global land use. Environmental Research Letters *16*, 125012. 10.1088/1748-9326/ac3c6d.
12. Stadelmann-Steffen, I., Eder, C., Harring, N., Spilker, G., and Katsanidou, A. (2021). A framework for social tipping in climate change mitigation: What we can learn about social tipping dynamics from the chlorofluorocarbons phase-out. Energy Research & Social Science *82*, 102307. https://doi.org/10.1016/j.erss.2021.102307.
13. Pascual, U., McElwee, P.D., Diamond, S.E., Ngo, H.T., Bai, X., Cheung, W.W.L., Lim, M., Steiner, N., Agard, J., Donatti, C.I., et al. (2022). Governing for Transformative Change across the Biodiversity-Climate-Society Nexus. BioScience *72*, 684-704. 10.1093/biosci/biac031.

14. Moore, F.C., Lacasse, K., Mach, K.J., Shin, Y.A., Gross, L.J., and Beckage, B. (2022). Determinants of emissions pathways in the coupled climate–social system. Nature. 10.1038/s41586-022-04423-8.
15. Bernstein, S., and Hoffmann, M. (2019). Climate politics, metaphors and the fractal carbon trap. Nature Climate Change *9*, 919-925.
16. Sterl, S., Hagemann, M., Fekete, H., Höhne, N., Cantzler, J., Ancygier, A., and Menon, S. (2017). Faster and cleaner 2: Kick-starting global decarbonization. Berlin, Germany: New Climate Institute, Climate Analytics, Ecofys, ClimateWorks Foundation.
17. Bhowmik, A.K., McCaffrey, M.S., Ruskey, A.M., Frischmann, C., and Gaffney, O. (2020). Powers of 10: seeking 'sweet spots' for rapid climate and sustainability actions between individual and global scales. Environmental Research Letters *15*, 094011. 10.1088/1748-9326/ab9ed0.
18. McCaffrey, M.S., and Boucher, J.L. (2022). Pedagogy of agency and action, powers of 10, and fractal entanglement: Radical means for rapid societal transformation toward survivability and justice. Energy Research & Social Science *90*, 102668. https://doi.org/10.1016/j.erss.2022.102668.
19. Climate Action Tracker (2019). Transformation points: Achieving the speed and scale required for full decarbonisation. Ecofys, New Climate Institute, Climate Analytics. climateactiontracker.org.
20. Smith, S.R., Christie, I., and Willis, R. (2020). Social tipping intervention strategies for rapid decarbonization need to consider how change happens. Proceedings of the National Academy of Sciences of the United States of America *117*, 10629-10630. 10.1073/pnas.1918465117.
21. Truelove, H.B., Carrico, A.R., Weber, E.U., Raimi, K.T., and Vandenbergh, M.P. (2014). Positive and negative spillover of pro-environmental behavior: An integrative review and theoretical framework. Global Environmental Change *29*, 127-138.
22. Heerma van Voss, B., and Rafaty, R. (2022). Sensitive intervention points in China's coal phaseout. Energy Policy *163*, 112797. 10.1016/j.enpol.2022.112797.
23. Höhne, N., Roy, J., Gaffney, O., Falk, J., Ribeiro, S.K., Levin, K., Smit, S., Nascimento, L., Hsu, A., and Mapes, B. (2021). A seismograph for measuring the transformation to net-zero greenhouse gas emissions: Discussion paper. New Climate Institute.
24. Systems Change Lab (2022). https://www.systemschangelab.org.
25. ClimateWatch (2022). Net Zero Tracker. https://www.climatewatchdata.org/net-zero-tracker.
26. Elliot, T. (2022). Socio-ecological contagion in Veganville. Ecol. Complex. *51*, 101015. 10.1016/j.ecocom.2022.101015.
27. Eker, S., Reese, G., and Obersteiner, M. (2019). Modelling the drivers of a widespread shift to sustainable diets. Nature Sustainability *2*, 725-735. 10.1038/s41893-019-0331-1.
28. Kitzmann, N.H., Romanczuk, P., Wunderling, N., and Donges, J.F. (2022). Detecting contagious spreading of urban innovations on the global city network. The European Physical Journal Special Topics. 10.1140/epjs/s11734-022-00470-4.

29. Kaaronen, R.O., and Strelkovskii, N. (2020). Cultural Evolution of Sustainable Behaviors: Pro-environmental Tipping Points in an Agent-Based Model. One Earth *2*, 85-97. 10.1016/j.oneear.2020.01.003.
30. Castilla-Rho, J.C., Rojas, R., Andersen, M.S., Holley, C., and Mariethoz, G. (2017). Social tipping points in global groundwater management. Nature Human Behaviour *1*, 640-649. 10.1038/s41562-017-0181-7.
31. Dunne, D., McSweeney, R., and Tandon, A. (2022). Tipping points: How could they shape the world's response to climate change? https://www.carbonbrief.org/tipping-points-how-could-they-shape-the-worlds-response-to-climate-change/.
32. Temper, L., Avila, S., Bene, D.D., Gobby, J., Kosoy, N., Billon, P.L., Martinez-Alier, J., Perkins, P., Roy, B., Scheidel, A., and Walter, M. (2020). Movements shaping climate futures: A systematic mapping of protests against fossil fuel and low-carbon energy projects. Environmental Research Letters *15*, 123004. 10.1088/1748-9326/abc197.
33. Axelrod, R., Daymude, J.J., and Forrest, S. (2021). Preventing extreme polarization of political attitudes. Proceedings of the National Academy of Sciences *118*.
34. Macy, M.W., Ma, M., Tabin, D.R., Gao, J., and Szymanski, B.K. (2021). Polarization and tipping points. Proceedings of the National Academy of Sciences *118*.
35. Rammelt, C.F., Gupta, J., Liverman, D., Scholtens, J., Ciobanu, D., Abrams, J.F., Bai, X., Gifford, L., Gordon, C., and Hurlbert, M. (2022). Impacts of meeting minimum access on critical earth systems amidst the Great Inequality. Nature Sustainability, 1-10.
36. Welsch, H. (2022). Do social norms trump rational choice in voluntary climate change mitigation? Multi-country evidence of social tipping points. Ecological Economics *200*, 107509. 10.1016/j.ecolecon.2022.107509.
37. Centola, D., Becker, J., Brackbill, D., and Baronchelli, A. (2018). Experimental evidence for tipping points in social convention. Science *360*, 1116-1119. doi:10.1126/science.aas8827.
38. Andreoni, J., Nikiforakis, N., and Siegenthaler, S. (2021). Predicting social tipping and norm change in controlled experiments. Proceedings of the National Academy of Sciences of the United States of America *118*, e2014893118. 10.1073/pnas.2014893118.
39. Ehret, S., Constantino, S.M., Weber, E.U., Efferson, C., and Vogt, S. (2022). Group identities can undermine social tipping after intervention. Nature Human Behaviour. 10.1038/s41562-022-01440-5.
40. Berger, J. (2021). Social tipping interventions can promote the diffusion or decay of sustainable consumption norms in the field. Evidence from a quasi-experimental intervention study. Sustainability *13*, 3529. 10.3390/su13063529.
41. Franzke, C.L., Ciullo, A., Gilmore, E.A., Matias, D.M., Nagabhatla, N., Orlov, A., Paterson, S.K., Scheffran, J., and Sillmann, J. (2022). Perspectives on tipping points in integrated models of the natural and human Earth system: cascading effects and telecoupling. Environmental Research Letters *17*, 015004.
42. Lenton, T.M., Benson, S., Smith, T., Ewer, T., Lanel, V., Petykowski, E., Powell, T.W., Abrams, J.F., Blomsma, F., and Sharpe, S. (2022).

Operationalising positive tipping points towards global sustainability. Global Sustainability *5*.
43. Meadows, D.H. (2008). Thinking in systems: A primer (chelsea green publishing).
44. Eker, S., Zimmermann, N., Carnohan, S., and Davies, M. (2017). Participatory system dynamics modelling for housing, energy and wellbeing interactions. Building Research & Information *46*, 738-754. 10.1080/09613218.2017.1362919.
45. Barbrook-Johnson, P., and Penn, A.S. (2022). Systems Mapping: How to build and use causal models of systems (Springer Nature).
46. Nyborg, K., Anderies, J.M., Dannenberg, A., Lindahl, T., Schill, C., Schlüter, M., Adger, W.N., Arrow, K.J., Barrett, S., and Carpenter, S. (2016). Social norms as solutions. Science *354*, 42-43.
47. Green, F. (2018). Anti-fossil fuel norms. Climatic Change *150*, 103-116. 10.1007/s10584-017-2134-6.
48. Schifeling, T., and Hoffman, A.J. (2017). Bill McKibben's Influence on U.S. Climate Change Discourse: Shifting Field-Level Debates Through Radical Flank Effects. Organization & Environment *32*, 213-233. 10.1177/1086026617744278.
49. Sabherwal, A., Ballew, M.T., van der Linden, S., Gustafson, A., Goldberg, M.H., Maibach, E.W., Kotcher, J.E., Swim, J.K., Rosenthal, S.A., and Leiserowitz, A. (2021). The Greta Thunberg Effect: Familiarity with Greta Thunberg predicts intentions to engage in climate activism in the United States. Journal of Applied Social Psychology *51*, 321-333. https://doi.org/10.1111/jasp.12737.
50. Wallis, H., and Loy, L.S. (2021). What drives pro-environmental activism of young people? A survey study on the Fridays For Future movement. Journal of Environmental Psychology *74*, 101581. https://doi.org/10.1016/j.jenvp.2021.101581.
51. Sunstein, C.R. (1996). Social norms and social roles. Colum. L. Rev. *96*, 903.
52. McCarthy, J. (2019). Most Americans Support Reducing Fossil Fuel Use. https://news.gallup.com/poll/248006/americans-support-reducing-fossil-fuel.aspx.
53. Aasen, M. (2017). The polarization of public concern about climate change in Norway. Climate Policy *17*, 213-230. 10.1080/14693062.2015.1094727.
54. Gustafson, A., Rosenthal, S.A., Ballew, M.T., Goldberg, M.H., Bergquist, P., Kotcher, J.E., Maibach, E.W., and Leiserowitz, A. (2019). The development of partisan polarization over the Green New Deal. Nature Climate Change *9*, 940-944. 10.1038/s41558-019-0621-7.
55. Brown, G., and Sovacool, B.K. (2017). The presidential politics of climate discourse: Energy frames, policy, and political tactics from the 2016 Primaries in the United States. Energy Policy *111*, 127-136. https://doi.org/10.1016/j.enpol.2017.09.019.
56. Bail, C.A., Argyle, L.P., Brown, T.W., Bumpus, J.P., Chen, H., Hunzaker, M.B.F., Lee, J., Mann, M., Merhout, F., and Volfovsky, A. (2018). Exposure to opposing views on social media can increase political polarization. Proceedings of the National Academy of Sciences *115*, 9216-9221. doi:10.1073/pnas.1804840115.

57. Mayer, A.P., and Smith, E.K. (2023). Multidimensional partisanship shapes climate policy support and behaviours. Nature Climate Change. 10.1038/s41558-022-01548-6.
58. Hielscher, S., Wittmayer, J.M., and Dańkowska, A. (2022). Social movements in energy transitions: The politics of fossil fuel energy pathways in the United Kingdom, the Netherlands and Poland. The Extractive Industries and Society *10*, 101073. https://doi.org/10.1016/j.exis.2022.101073.
59. Brulle, R.J., Aronczyk, M., and Carmichael, J. (2020). Corporate promotion and climate change: an analysis of key variables affecting advertising spending by major oil corporations, 1986–2015. Climatic Change *159*, 87-101. 10.1007/s10584-019-02582-8.
60. Farrell, J. (2016). Corporate funding and ideological polarization about climate change. Proceedings of the National Academy of Sciences *113*, 92-97. doi:10.1073/pnas.1509433112.
61. Hoffmann, R., Muttarak, R., Peisker, J., and Stanig, P. (2022). Climate change experiences raise environmental concerns and promote Green voting. Nature Climate Change *12*, 148-155. 10.1038/s41558-021-01263-8.
62. Drake, H.F., and Henderson, G. (2022). A defense of usable climate mitigation science: how science can contribute to social movements. Climatic Change *172*, 1-18.
63. IEA (2022). World Energy Investment. International Energy Agency.
64. Atanasova, C., and Schwartz, E.S. (2019). Stranded fossil fuel reserves and firm value. National Bureau of Economic Research.
65. Battiston, S., Monasterolo, I., Riahi, K., and van Ruijven, B.J. (2021). Accounting for finance is key for climate mitigation pathways. Science *372*, 918-920. doi:10.1126/science.abf3877.
66. Battiston, S., Mandel, A., Monasterolo, I., Schütze, F., and Visentin, G. (2017). A climate stress-test of the financial system. Nature Climate Change *7*, 283-288.
67. Victor, D.G., Lumkowsky, M., and Dannenberg, A. (2022). Determining the credibility of commitments in international climate policy. Nature Climate Change *12*, 793-800. 10.1038/s41558-022-01454-x.
68. Dietz, S., Bowen, A., Dixon, C., and Gradwell, P. (2016). 'Climate value at risk' of global financial assets. Nature Climate Change *6*, 676-679. 10.1038/nclimate2972.
69. Monasterolo, I., and de Angelis, L. (2020). Blind to carbon risk? An analysis of stock market reaction to the Paris Agreement. Ecological Economics *170*, 106571. https://doi.org/10.1016/j.ecolecon.2019.106571.
70. Iacobuta, G., Dubash, N.K., Upadhyaya, P., Deribe, M., and Höhne, N. (2018). National climate change mitigation legislation, strategy and targets: a global update. Climate Policy *18*, 1114-1132. 10.1080/14693062.2018.1489772.
71. Erickson, P., Lazarus, M., and Piggot, G. (2018). Limiting fossil fuel production as the next big step in climate policy. Nature Climate Change *8*, 1037-1043. 10.1038/s41558-018-0337-0.
72. IEA (2022). World Energy Balances. International Energy Agency. https://www.iea.org/data-and-statistics/data-product/world-energy-balances.
73. Gaulin, N., and Le Billon, P. (2020). Climate change and fossil fuel production cuts: assessing global supply-side constraints and policy implications. Climate Policy *20*, 888-901. 10.1080/14693062.2020.1725409.

74. Carattini, S., and Löschel, A. (2021). Managing momentum in climate negotiations*. Environmental Research Letters *16*, 051001. 10.1088/1748-9326/abf58d.
75. Moore, F.C., Obradovich, N., Lehner, F., and Baylis, P. (2019). Rapidly declining remarkability of temperature anomalies may obscure public perception of climate change. Proceedings of the National Academy of Sciences *116*, 4905-4910. doi:10.1073/pnas.1816541116.
76. Kraft-Todd, G.T., Bollinger, B., Gillingham, K., Lamp, S., and Rand, D.G. (2018). Credibility-enhancing displays promote the provision of non-normative public goods. Nature *563*, 245-248. 10.1038/s41586-018-0647-4.
77. Oreskovic, L., and Gupta, R. (2022). Enabling Sustainable Lifestyles in New Urban Areas: Evaluation of an Eco-Development Case Study in the UK. Sustainability *14*, 4143. 10.3390/su14074143.
78. Kraus, S., and Koch, N. (2021). Provisional COVID-19 infrastructure induces large, rapid increases in cycling. Proceedings of the National Academy of Sciences *118*, e2024399118.
79. Granoff, I., Hogarth, J.R., and Miller, A. (2016). Nested barriers to low-carbon infrastructure investment. Nature Climate Change *6*, 1065-1071. 10.1038/nclimate3142.
80. Dodds, P.S., and Watts, D.J. (2005). A generalized model of social and biological contagion. Journal of Theoretical Biology *232*, 587-604. https://doi.org/10.1016/j.jtbi.2004.09.006.
81. Dobbin, F., Simmons, B., and Garrett, G. (2007). The global diffusion of public policies: Social construction, coercion, competition, or learning? Annu. Rev. Sociol. *33*, 449-472.
82. Pamlin, D. (2021). 21st century climate alignment assessment: Identifying 1.5°C compatible innovations in the 4th industrial revolution that can deliver what is needed to allow 11 billion people live flourishing lives. Mission Innovation Net-Zero Compatibility Innovations Initiative and Research Institutes Sweden (RISE).
83. Eker, S., Garcia, D., Valin, H., and van Ruijven, B. (2021). Using social media audience data to analyse the drivers of low-carbon diets. Environmental Research Letters.
84. Inglehart, R., Haerpfer, C., Moreno, A., Welzel, C., Kizilova, K., Diez-Medrano J., M. Lagos, P. Norris, E. Ponarin & B. Puranen et al. (2020). World Values Survey: All Rounds – Country-Pooled Datafile. In J.S.I.W. Secretariat, ed.
85. van Beek, L., Hajer, M., Pelzer, P., van Vuuren, D., and Cassen, C. (2020). Anticipating futures through models: the rise of Integrated Assessment Modelling in the climate science-policy interface since 1970. Global Environmental Change *65*, 102191.
86. Beckage, B., Lacasse, K., Winter, J.M., Gross, L.J., Fefferman, N., Hoffman, F.M., Metcalf, S.S., Franck, T., Carr, E., and Zia, A. (2020). The Earth has humans, so why don't our climate models? Climatic Change *163*, 181-188.
87. Trutnevyte, E., Hirt, L.F., Bauer, N., Cherp, A., Hawkes, A., Edelenbosch, O.Y., Pedde, S., and van Vuuren, D.P. (2019). Societal transformations in models for energy and climate policy: the ambitious next step. One Earth *1*, 423-433.

88. Farmer, J.D., Hepburn, C., Mealy, P., and Teytelboym, A. (2015). A Third Wave in the Economics of Climate Change. Environmental and Resource Economics *62*, 329-357. 10.1007/s10640-015-9965-2.
89. Lamperti, F., Dosi, G., Napoletano, M., Roventini, A., and Sapio, A. (2018). Faraway, So Close: Coupled Climate and Economic Dynamics in an Agent-based Integrated Assessment Model. Ecological Economics *150*, 315-339. https://doi.org/10.1016/j.ecolecon.2018.03.023.
90. Randers, J., Rockström, J., Stoknes, P.-E., Goluke, U., Collste, D., Cornell, S.E., and Donges, J. (2019). Achieving the 17 Sustainable Development Goals within 9 planetary boundaries. Global Sustainability *2*, e24, e24. 10.1017/sus.2019.22.
91. Moallemi, E.A., Eker, S., Gao, L., Hadjikakou, M., Liu, Q., Kwakkel, J., Reed, P.M., Obersteiner, M., Guo, Z., and Bryan, B.A. (2022). Early systems change necessary for catalyzing long-term sustainability in a post-2030 agenda. One Earth *5*, 792-811. https://doi.org/10.1016/j.oneear.2022.06.003.
92. Beckage, B., Moore, F.C., and Lacasse, K. (2022). Incorporating human behaviour into Earth system modelling. Nature Human Behaviour *6*, 1493-1502. 10.1038/s41562-022-01478-5.
93. Granovetter, M. (1978). Threshold models of collective behavior. American journal of sociology *83*, 1420-1443.
94. Wiedermann, M., Smith, E.K., Heitzig, J., and Donges, J.F. (2020). A network-based microfoundation of Granovetter's threshold model for social tipping. Sci. Rep. *10*, 11202. 10.1038/s41598-020-67102-6.
95. Lavoie, M., and Zezza, G. (2012). The stock-flow consistent approach: selected writings of Wynne Godley (Palgrave Macmillan).
96. Rahmandad, H., and Sterman, J. (2008). Heterogeneity and network structure in the dynamics of diffusion: Comparing agent-based and differential equation models. Management Science *54*, 998-1014.
97. Köhler, J., De Haan, F., Holtz, G., Kubeczko, K., Moallemi, E., Papachristos, G., and Chappin, E. (2018). Modelling sustainability transitions: An assessment of approaches and challenges. Journal of Artificial Societies and Social Simulation *21*.
98. Niamir, L., Ivanova, O., and Filatova, T. (2020). Economy-wide impacts of behavioral climate change mitigation: Linking agent-based and computable general equilibrium models. Environmental Modelling & Software *134*, 104839. https://doi.org/10.1016/j.envsoft.2020.104839.
99. Sterman, J. (2018). System dynamics at sixty: the path forward. System Dynamics Review *34*, 5-47. https://doi.org/10.1002/sdr.1601.

## Acknowledgements

We acknowledge financial support from Horizon Europe (WorldTrans, grant agreement ID: 101081661), European Research Council (iDODDLE, grant agreement ID: 101003083) and Mission Innovation Net-Zero Compatibility Initiative. We thank Franziska Gaupp, Jean-Francois Mercure, Raya Muttarak, E. Keith Smith, Dennis Pamlin, and Ilona M. Otto for their valuable participation in the expert workshop.

## Contributions

S.E. and C.W. designed and conducted the research. N.H., M.S.M., I.M, L.N., and C.Z. contributed with data and disciplinary ideas. S.E. and C.W. drafted the text and figures. All authors reviewed and edited the paper.

## Competing interests

The authors declare no competing interests.